\documentclass[11pt]{iopart}
\pdfoutput=1
\usepackage{geometry}                
\usepackage{graphicx}
\usepackage{amssymb}
\usepackage{epstopdf}
\usepackage[T1]{fontenc}
\usepackage[utf8]{inputenc}
\usepackage{float}

\newcommand{\mK}{\mathcal{K}}
\newcommand{\mS}{\mathcal{S}}
\newcommand{\mQ}{\mathcal{Q}}

\usepackage{alphalph}
\usepackage{bigfoot}
\DeclareNewFootnote[para]{B}[alph]
\DeclareNewFootnote{default}
\usepackage{perpage}
\MakePerPage{footnoteB}

\begin{document}
\title{Criticality of mostly informative samples: A Bayesian model selection approach}

\author{Ariel Haimovici}
\address{Departamento de F\'{i}sica, Facultad de Ciencias Exactas y Naturales, Universidad de Buenos Aires, Buenos Aires, Argentina}
\author{Matteo Marsili}
\address{The Abdus Salam International Centre for Theoretical Physics, Strada Costiera 11, 34151 Trieste, Italy}

\maketitle

\begin{abstract}
We discuss a Bayesian model selection approach to high dimensional data in the deep under sampling regime. The data is based on a representation of the possible discrete states $s$, as defined by the observer, and it consists of $M$ observations of the state. This approach shows that, for a given sample size $M$, not all states observed in the sample can be distinguished. Rather, only a partition of the sampled states $s$ can be resolved. Such partition defines an {\em emergent} classification $q_s$ of the states that becomes finer and finer as the sample size increases, through a process of {\em symmetry breaking} between states. 
This allows us to distinguish between the {\em resolution} of a given representation of the observer defined states $s$, which is given by the entropy of $s$, and its {\em relevance} which is defined by the entropy of the partition $q_s$. Relevance has a non-monotonic dependence on resolution, for a given sample size. 
In addition, we characterise most relevant samples and we show that they exhibit power law frequency distributions, generally taken as signatures of ``criticality''. This suggests that ``criticality'' reflects the relevance of a given representation of the states of a complex system, and does not necessarily require a specific mechanism of self-organisation to a critical point.
\end{abstract}

\section{Introduction}
In the study of complex systems -- such as the brain, cells or our economies -- we face conceptual issues of a novel type, because the systems studied involve many variables, many of which are unknown. In addition, their behaviour is not constrained by well established laws, as in physics. 
In such high dimensional inference problems one is hardly ever sampling correctly an underlying probability distribution, even with huge data sets. In order to evade the deep under-sampling domain, we implicitly or explicitly resort to dimensionality reduction schemes, where the data is projected into a low-dimensional space where statistics can provide accurate conclusions. Yet, in this process, the data processing inequality \cite{cover91} tells us that we inevitably loose relevant information on the system's ``laws of motion''. So understanding which are the relevant variables is crucial in order to limit information losses. This requires guiding principles for the choice of dimensional reduction schemes, or for measuring the relevance of a given set of variables.

Recently Ref. \cite{marsili13} suggested that the entropy of the frequency of observations (see later) can be used as a measure of relevance of a given representation of the data. This allows one to characterise {\em most informative samples} as those that maximise this measure, at a given resolution and for a given sample size. Remarkably, one finds that most informative samples, in the under-sampling regime, have a power law frequency distribution \cite{marsili13}.

This finding sheds light on the widespread observation of ``criticality''  (i.e. power law frequency/size distributions) in empirical data \cite{clauset09} ranging from language \cite{zipf32}, statistics of natural images \cite{ruderman94}, neural activity \cite{eguiluz05,schneidman06}, city size distribution \cite{gabaix99}, to name just a few cases. In brief, this strongly suggests that the observed power laws usually associated with ``criticality'' arise as a consequence of our choice of relevant variables and that they do not necessarily require hidden mechanisms of self-organisation to a critical point \cite{bak96}. Besides the academic interest of such an interpretation of ``criticality'', its implication for data analysis are far reaching because the proposed measure of relevance can be used as a universal guiding principle in the search of optimal dimensional reduction schemes (e.g. data clustering) or for the identification of relevant variables (e.g. keywords in texts, relevant amino acids in proteins) \cite{marsili13}. 

The purpose of this paper is to ground the finding of Ref. \cite{marsili13} in a model selection Bayesian framework, thereby clarifying its information theoretic basis. In brief, within this approach, we shall see statistical models of the data emerge from a process of symmetry breaking between data points in the sample\footnote{In what follows, a sample is a sequence of data points, each of which belong to a set of possible outcomes, which are defined {\em a priori}.}, acquiring more and more details as the size of the sample increases. In this way, model selection informs us on what resolution in the space of outcomes is justified by the data. In order for different outcomes to be assigned different probabilities, the frequency with which they occur in the sample must be sufficiently different. Formally, this identifies an {\em optimal partition} which distinguishes outcomes that occur with different probabilities. The entropy of the size of the partitions provides a measure of the number of outcomes that can be distinguished in the sample (or of the number of parameters that can be estimated from the samples) and hence a measure of relevance. In what follows, for the sake of simplicity, we shall define and refer to this measure as \emph{relevance}.

The next section introduces the generic problem we deal with and discusses model selection. Simple examples are presented to provide the main intuition. 
We shall first show that, barring atypical cases, an upper bound to the relevance is given by partitions in frequency classes. Next we shall see that most informative samples are characterised by power law frequency distributions. 
This will be followed by an application to two different examples of real data sets. The results suggest that the entropy of the frequency as suggested in Ref. \cite{marsili13}, can be used in place of the entropy of the optimal partition, which is computationally more demanding, as a measure of relevance. A final discussion will close the paper.

\section{The problem}

Let $\hat s=(s^{(1)},\ldots,s^{(M)})$ be a dataset of $M$ observations of the state $s$ of a system. 
Here $s^{(i)}$ is a discrete variable, that we can think of as the label of the cluster to which the $i^{\rm th}$ observation belongs, or the configuration $s=(s_1,\ldots,s_n)$ of a system of $n$ discrete degrees of freedom (e.g. the amino acid sequence of a protein domain). The number of possible different states $s$ is much larger than $M$ and it may even be unknown. We restrict attention to the case where $s^{(i)}$ can be thought of as outcomes of independent experiments, carried out in the same conditions.

The general question of interest is to infer the laws governing the system, from the data. This can be formalised by assuming that the data can be thought of as $M$ i.i.d. draws from a generative model $P\{s^{(i)}=s\}=p_s$, where the function $p_s$ should encode the property of the system and the functions it performs. The basic problem then becomes that of inferring the generative model.

\subsection{Resolution and relevance}

Reference \cite{marsili13} has shown that, if we think of each sample $s^{(i)}$ as a realisation of an optimisation problem of a function $U(s,\bar s)$ over an enlarged set of variables that includes also unknown variables ($\bar s$), then the frequency 
\[
k_s =\sum_{i=1}^M \delta_{s^{(i)},s}
\]
with which a given observation $s$ occurs in the sample provides a noisy estimate of that part $u_s=E_{\bar s}[U(s,\bar s)]$ of the function that is being optimised. Hence the relevance of the particular choice of the variables $s$, among all those that enter $U$, is reflected in the statistics of the frequency $k_s$ of states $s$. Ref. \cite{marsili13} argues that a quantitative measure of \emph{relevance}, in information theoretic terms, is given by
\begin{equation}
\label{eqHk}
\hat H[K]=-\sum_{k} \frac{k m_k}{M}\log  \frac{km_k}{M}, 
\end{equation}
where 
\[
m_k=\sum_s\delta_{k_s,k}
\]
is the number of states that occur $k$ times in the sample $\hat s$. Notice that $\hat H[K]$ is the entropy of the random variable $K_i=k_{s^{(i)}}$ for a randomly chosen point $s^{(i)}$ of the sample. This is different from the entropy of the state $s$ itself \footnote{Again we use uppercase for random variables defined on the space of the points in the sample $\hat s$. Also we assume maximum likelihood estimates of the probability $P\{S=s\}=k_s/M$.}
\begin{equation}
\label{eqHs}
\hat H[S]=-\sum_s\frac{k_s}{M}\log\frac{k_s}{M}=-\sum_{k} \frac{k m_k}{M}\log  \frac{k}{M}.
\end{equation}
Intuitively, this measures the {\em resolution} of the description based on the variable $s$. Indeed a more detailed definition of the state $s$ of the system likely results in a higher {\em resolution} (i.e. a larger value of) $\hat H[S]$ but not necessarily in a higher {\em relevance} $\hat H[K]$. 

\subsection{Learning the generative model}

Given a generative model $\mathcal{M}=p_s$, the likelihood of $\hat s$ is defined as:
\begin{equation}
\label{ }
P(\hat s|\mathcal{M})=\prod_{i=1}^M p_{s^{(i)}}=\prod_s p_s^{k_s},~~~~k_s=\sum_{i=1}^M \delta_{s,s^{(i)}}.
\end{equation}
The frequentist approach estimates the best model as the one  that maximises the likelihood. This results in equating probabilities with frequencies: $\hat p_s=k_s/M$. The Bayesian approach, instead, invokes Bayes rule to turn the likelihood into a (posterior) distribution over the parameters $\vec p$ of the model. This requires identifying a {\em prior} distribution $P_0(\vec{p})$ that reflects our ignorance on $\vec p$ before seeing the data. A minimal requirement is that $P_0(\vec p)$ should be a symmetric function of the $p_s$'s. Dirichelet priors 

\begin{equation}
\label{ }
P_0(\vec p)=\Gamma\left(\sum_{s} a_{s}\right)\prod_{s}\frac{p_{s}^{a_{s}-1}}{\Gamma(a_{s})}\delta\left(\sum_{s} p_{s}-1\right)
\end{equation}
are a mathematically convenient choice, and ignorance requires by symmetry that $a_s=a$ is independent of $s$. The posterior is easily computed:
\begin{equation}
\label{ }
P_1(\vec p)=\Gamma\left[\sum_{s} (k_{s}+a)\right]\prod_{s}\frac{p_{s}^{k_{s}+a-1}}{\Gamma(k_{s}+a)}\delta\left(\sum_{s} p_{s}-1\right)
\end{equation}

This allows us to give a Bayesian estimate of the probabilities 
\begin{equation}
\label{ }
\langle p_s\rangle_1=\int\!d\vec p p_s P_1(\vec p)=\frac{k_s+a}{M+aS}
\end{equation}
where $S$ is the number of states. When $M\gg aS$ this converges to the frequentist estimate $k_s/M$, reminding us that in the presence of a large enough data set, the choice of the prior does not matter.

There are a number of problematic issues with this procedure:
\begin{enumerate}
  \item The set of possible states and their number $S$ should be known in advance. This is not always the case.
  \item The estimate of the entropy $H[S]=-\sum_s \langle p_s\log p_s\rangle_1$ is strongly affected by the prior and it converges slowly to its true value, as shown in Ref. \cite{nemenman01}. 
  \item The model assumes a different parameter for each state that occurs in the sample. A posteriori, this assumption is not justified as there is nothing that can be learned from the data on how the probabilities of two states that are seen the same number of times differ. Indeed, the posterior estimate of these probability depends on the frequency $k_s$ and is exactly the same for two states $s,s'$ that occur the same number of times $k_s=k_{s'}$.
\end{enumerate}

In particular, the last point suggests that we are in a clear case of over-fitting and indeed this model does not survive a model selection test, as we shall see in what follows.

\section{Model selection}

The key issue is that the definition of states $s$ is made by the observer, not by the system. 
If the distinction between $s$ and $s'$ is totally spurious, we expect that the data will not distinguish between the two states, i.e. $k_s \approx k_{s'}$. Conversely, if two states are seen the same number of times, there is no reason to assume that they have a different probability. In terms of inference, we are not allowed to think that $p_s \neq p_{s'}$ unless we have sufficient evidence.

\subsection{An illustrative case: two states}

Let there be only two states $s=0,1$ and assume there are $M$ observations, $k=k_1$ with $s=1$ and $M-k$ with $s=0$. There are two possibilities: one is that the two states are actually the same, i.e. that the underlying distribution has $p_0=p_1=1/2$, the other that the states are different, i.e. $p_1=p=1-p_0$. 
These correspond to different models that we can identify with different partitions of  states and the associated probabilities. So the first case corresponds to a model $\mathcal{M}_0=[(\{0,1\},1/2)]$ where the two states are symmetric, whereas the second to a model $\mathcal{M}_1=[(\{0\},1-p),(\{1\},p)]$. 
Clearly $P\{\hat s|\mathcal{M}_0\}=2^{-M}$ whereas for $\mathcal{M}_1$ the likelihood $P\{\hat s|\mathcal{M}_1\}$ can be obtained by integrating the likelihood over the prior distribution of the parameter $p$, for which again we take a Dirichelet form. Hence
\begin{equation}
\label{ }
P\{\hat s|\mathcal{M}_1\}=\frac{\Gamma(2a)\Gamma(k+a)\Gamma(M-k+a)}{\Gamma(a)^2\Gamma(M+2a)}.
\end{equation}
In order to compare the two models, we invoke Bayes rule and compute the posterior probability 
\[
P(\mathcal{M}_i|\hat{s})=\frac{P(\hat{s}|\mathcal{M}_i)P_0(\mathcal{M}_i)}
{\sum_j P(\hat{s}|\mathcal{M}_j)P_0(\mathcal{M}_j)}=\frac{P(\hat{s}|\mathcal{M}_i)P_0(\mathcal{M}_i)}
{P(\hat{s})}
\]
where $P_0(\mathcal{M}_i)$ is the prior probability of model $i$. For the sake of simplicity, we're going to assume that all models are {\em a priori} equally likely\footnote{By Occam's razor, one would be tempted to prefer simpler models, i.e. those with fewer parameters. Yet Occam's razor already arises from the integration over the parameters implied by Bayes rule, without the need to introduce it {\em ad hoc}.}. 
So the most probable model is the one with the highest likelihood $P\{\hat s|\mathcal{M}\}$. In the present case, it is easy to check that, for $M\gg 1$, in the representative case of a uniform prior ($a=1$) we have that  as long as
\[
\left|\frac{k}{M}-\frac{1}{2}\right|<\sqrt{\frac{\log(2M/\pi)}{2M}}
\]
the symmetric model $\mathcal{M}_0$ should be preferred.

\begin{figure}
\centering
\includegraphics[width=0.6\textwidth]{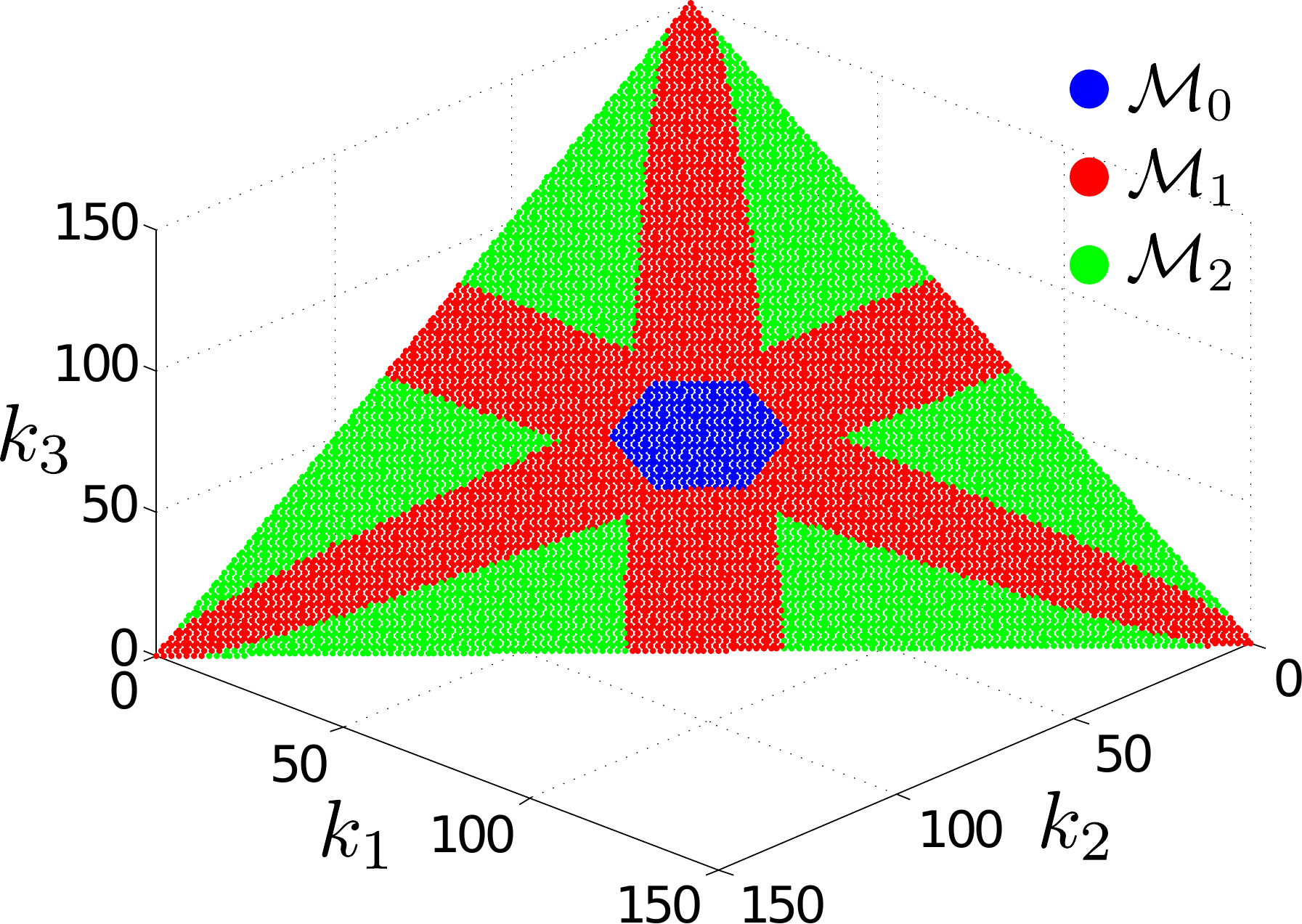}
\caption{Model selection in a three state system with $M=150$ observations. $k_1$, $k_2$ and $k_3$ are the number of observations of each state. The coloured surface shows the preferred model in terms of the likelihood $P(\hat{s}|\mathcal{M}_i)$. $\mathcal{M}_0$ is the model with no parameters ($p_i=1/3, \forall i$), $\mathcal{M}_1$ is the one with one parameter ($p_i=p, ~p_{j}=p_k=(1-p)/2$), and $\mathcal{M}_2$ is the one with two parameters ($p_i=p,~ p_j=q,p_k=(1-(p+q))$).}
\label{three_states}
\end{figure}

Figure \ref{three_states} shows an extension for the 3-states case. Here the possible models are $\mathcal{M}_0$ with no parameters (each state has $p=1/3$), $\mathcal{M}_{1,i}$ where two out of the three states have the same probability ($p_i=p$ and $p_s=(1-p)/2$ for $i=1,2$ or $3$), and $\mathcal{M}_2$ where all states have a different probability. If the frequencies are close enough, the states should not be distinguished and the model with no parameters should be preferred (blue surface in \ref{three_states}). Conversely the red surface reflect the cases where two states should not be distinguished from each other, and the green shows the case were the three states should be distinguished.

\subsection{The general case}

The argument above suggests that, in the general case, for each pair of states $s$ and $s'$ their probability should be the same, unless they occur in the data a sufficiently different number of times. If $k_s\approx k_{s'}$ instead, they should be assigned the same probability, i.e. the symmetry between states $s$ and $s'$ should not be broken.

Conversely, imagine the situation where the distinction between states $s$ and $s'$ is completely arbitrary, with no relation with the internal states of the system under study. Complete ignorance of the system about the distinction between states $s$ and $s'$ means that the probability distribution restricted to only these two states must be the one of maximal entropy, i.e. that $p_s=p_{s'}$. 

We remind again that the definition of states $s$ is made by the observer, not by the system. If it distinguishes effectively different internal states of the system, then this definition is relevant and meaningful, otherwise it is not. One way to turn this observation into a quantitative criterium is to extend the model selection argument above.

Given the set $\mathcal{S}$ of states $s$ that are seen (with multiplicity $k_{s}>0$), then a generic model $\mathcal{M}=[\mathcal{Q},\vec \mu]$ is one where different states are divided into a partition 
\[
\mathcal{Q}=(Q_1,Q_2,\ldots,Q_N),\qquad \bigcup_{q=1}^N Q_q=\mathcal{S}
\]
of a number $N$ of disjoint sets, and each state in the $q^{\rm th}$ subset of the partition ($s\in Q_q$) has the same probability\footnote{All quantities $N$, $m_q$, $Q_q$ $\mu_q$ depend on the model $\mathcal{M}$. We omit this dependence for the sake of simplifying formulas.} $\mu_q$. If $m_q=|Q_q|$ is the number of states in subset $Q_q$, then $\mu_q$ satisfies the normalisation
\begin{equation}
\label{normal}
\sum_{q} m_q\mu_q=1.
\end{equation}

Any possible partition corresponds to a different model, going from the one where each state is in the same subset ($s\in Q_1,\forall s$), to the one where each state is in a different subset ($s\in Q_{s}, \forall s$).
It is possible to consider more general structure that also includes yet not seen states (i.e. states with $k_s=0$). 
We shall see below that these are less likely than those considered here.
Each partition $\mathcal{Q}$ identifies a different model $\mathcal{M}$. This is why we shall use the partition $\mathcal{Q}$ to refer to the model that is based on that partition.

It is straightforward to compute the likelihood of each model:
\begin{equation}
\label{ }
P\{\hat s|\mathcal{Q}\}=\int\!d\vec \mu \prod_{q}\mu_q^{K_q}P_0^{(\mathcal{Q})}(\vec \mu),\qquad K_q=\sum_{s\in Q_q}k_{s}
\end{equation}
where the prior $P_0^{(\mathcal{Q})}$ contains the constraint Eq. (\ref{normal}).
We take again conjugate (Dirichelet) priors
\begin{equation}
\label{ }
P_0^{(\mathcal{Q})}(\vec \mu)=\Gamma(aN)\prod_{q}\frac{m_q^a}{\Gamma(a)}\mu_q^{a-1}\delta\left(\sum_{q\in\mathcal{Q}} m_q\mu_q-1\right)
\end{equation}
where $N$ is the number of partitions in $\mathcal{Q}$, i.e of parameters in $\mathcal{Q}$. Then
\begin{equation}\label{likelihood}
\log P\{\hat s|\mathcal{Q}\}=\sum_{q}\left[\log\frac{\Gamma(K_q+a)}{\Gamma(a)}-K_q\log m_q\right]-\log\frac{\Gamma(M+aN)}{\Gamma(aN)}
\end{equation}

The posterior distribution, under model $\mathcal{Q}$ is
\begin{equation}
\label{P1}
P_1^{(\mathcal{Q})}(\vec \mu|\hat s)=\Gamma(M+aN)\prod_{q}\frac{m_q^{K_q+a}}{\Gamma(K_q+a)}\mu_q^{K_q+a-1}\delta\left(\sum_{q} m_q\mu_q-1\right)
\end{equation}	

The expected value of $p_{s}$ for $s\in Q_q$ is
\begin{equation}
\label{ps_post}
\left\langle p_{s}|\mathcal{Q}\right\rangle_{1}=\frac{1}{m_q}\frac{K_q+a}{M+aN}, \qquad \forall s\in Q_q
\end{equation}
where $\langle \ldots|\mathcal{Q}\rangle_{1}$ indicates expected values over the posterior distribution Eq. (\ref{P1}).
The expected value of the entropy $H[S]=-\sum_{s}p_{s}\log p_{s}$ is given by
\begin{equation}
\label{Hs}
\left\langle H[S]|\mathcal{Q}\right\rangle_{1}=-\sum_{q}\frac{K_q+a}{M+aN}\left[\psi(K_q+a+1)-\log m_q-\psi(M+aN+1)\right]
\end{equation}
where $\psi(z)=\frac{d\log \Gamma(z)}{dz}$ is the digamma function.


Assuming that all models are {\em a priori} equally likely, $P\{\hat s|\mathcal{Q}\}$ is also proportional to the posterior probability $P\{\mathcal{Q}|\hat s\}$ of model $\mathcal{Q}$ given the data. Therefore the optimal model is given by\footnote{A fully Bayesian approach would entail considering all possible partitions $\mathcal{Q}$ with their probability $P\{\mathcal{Q}|\hat s\}$. Here we depart from this approach and focus on the most likely partition.}
\begin{equation}
\label{ }
\mathcal{Q}^*={\rm arg}\max_{\mathcal{Q}} P\{\hat s|\mathcal{Q}\}.
\end{equation}

The  partition $\mathcal{Q}^*$ identifies an emergent description of the system in terms of effective states $q$, that we shall call $q$-states. This is the statistical description that can be resolved on the basis of the dataset $\hat s$. The states $s\in Q_q^*$ in the same partition $q$ cannot be distinguished one from the other, hence they all correspond to the same $q$-state.  The variable $q$ is associated to a distribution $p_q=m_q\mu_q$, which is the probability to observe the $q$-state. The entropy of this distribution $H[Q]=-\sum_q p_q\log p_q$ provides a quantitative measure of the amount of information that the data provides on the generative model. It's expected value on the posterior distribution Eq. (\ref{P1})
\begin{eqnarray}
\left\langle H[Q]|\mathcal{Q}\right\rangle_{1} & = & -\sum_{q}\frac{K_q+a}{M+aN}\left[\psi(K_q+a+1)-\psi(M+aN+1)\right] \\
 & = & \left\langle H[S]|\mathcal{Q}\right\rangle_{1}-\sum_{q}\frac{K_q+a}{M+aN}\log m_q \label{HQdiff}
\end{eqnarray}
is what we shall call {\em relevance}. Indeed, this is a measure of the relevance of the original description based on the states $s$. Eq. (\ref{HQdiff}) shows that $\left\langle H[Q]|\mathcal{Q}\right\rangle_{1} \le \left\langle H[S]|\mathcal{Q}\right\rangle_{1}$ with equality if and only if all partitions $Q_q$ contain only one state ($m_q=1~\forall q$).
The next section illustrates the behaviour of this measure in some specific examples. 
Before doing that, it is instructive to discuss the issue of unsampled states and two special cases, to make contact with the results of Ref. \cite{marsili13}.

\subsection{Unsampled states}
\label{sec:unsampled}
In many instances, the sample contains only a partial coverage of the set of possible states. There are two ways in which not yet sampled states could be included in one of the partitions $\mQ$ discussed above. Either adding them to one or more of the sets $Q_q$ or augmenting the partition with a set $Q_0$ that includes all states with $k_s=0$. In the first case, the partition $\mQ$ changes into one which is identical on all sets $Q_{q'}$ with $q'\neq q$ and with $Q_q\to Q_q'=Q_q\bigcup Q_0$, where $Q_0$ is the set of unseen states. Since $k_s=0$ for $s\in Q_0$, the count $K_q$ does not change, and the change in the likelihood is given by $-K_q\log(1+m_0/m_q)$, where $m_0=|Q_0|$ is the number of states $s\in Q_0$. Since the change in the likelihood is negative, the optimal partition $\mQ^*$ does not include not yet sampled states.

The change in the likelihood when the unseen states are added to the partition in a new set, $\mQ\to \mQ_{+0}=(\mQ,Q_0)$ can also be easily computed. The first two terms in Eq. (\ref{likelihood}) do not change, as $K_0=0$, so the only difference is due to the fact that the number of sets increases by one: $N\to N+1$. Hence the change in the likelihood 
\begin{equation}\label{dlikelihood}
\log \frac{P\{\hat s|\mathcal{Q}_{+0}\}}{P\{\hat s|\mathcal{Q}\}}=
-\log\frac{\Gamma(M+aN+a)\Gamma(aN)}{\Gamma(M+aN)\Gamma(aN+a)}
\end{equation}
is again negative.
Hence models based on  partitions that include unseen states are dominated by those discussed above, if they are considered equally likely {\em a priori}. 

Yet, if one expects that the sample contains only a partial coverage of the set of possible states, the uniform prior hypothesis needs to be revised. Therefore
\begin{equation}
\label{dpost}
\log \frac{P\{\mathcal{Q}_{+0}|\hat s\}}{P\{\mathcal{Q}|\hat s\}}=\Delta_0
-\sum_{k=0}^{M-1}\log\left(1+\frac{a}{k+aN}\right)
\end{equation}
where $\Delta_0=\log\frac{P_0\{\mathcal{Q}_{+0}\}}{P_0\{\mathcal{Q}\}}$ encodes the {\em a priori} likelihood that states $s$ that are not present in the sample $\hat s$ exist. Notice that the second term in Eq. (\ref{dpost}) increases with $M$ (as $a\log (1+M/(aN))$ for $M,N\gg 1$). Hence for a given $\Delta_0$, we expect the model $\mQ$ to become preferable to $\mQ_{+0}$ as $M$ grows large. When instead the model $\mQ_{+0}$ is the optimal, this approach also gives an estimate of the discovery probability 
\begin{equation}
\label{ }
p_0=\frac{a}{M+aN+a}
\end{equation}
which is an intense subject of research in statistical learning\footnote{This discussion relates to the wider field of non-parametric Bayesian statistics which discusses models that reproduce sampling processes. For a general introduction, the reader is referred to \cite{OrbanzTeh2010}. A model of the sampling process based on our approach departs from this literature in that non-parametric Bayesian models such as the Dirichelet's process are based on a single partition ($\mathcal{S}$ in this case) whereas we consider selecting the optimal partition for each $M$. Further discussion of this issue would bring us too far from the main aim of the present paper and will be dealt with elsewhere.}, since the work of Good and Turing \cite{goodturing}.

%

\subsection{Special cases}

For the model based on the atomic partition $\mathcal{S}$, where each subset contains one state $Q_s=\{s\}$ 
\begin{eqnarray}
\log P\{\hat s|\mathcal{S}\} & = & \sum_{s}\log\frac{\Gamma(k_{s}+a)}{\Gamma(a)}-\log\frac{\Gamma(M+aN_s)}{\Gamma(aN_s)} 
\\
 & = & \sum_k m_k\log\frac{\Gamma(k+a)}{\Gamma(a)}-\log\frac{\Gamma(M+aN_s)}{\Gamma(aN_s)}
\end{eqnarray}
where $N_s=|\mathcal{S}|$ is the number of different states $s$ that occur in the sample $\hat s$. Note that $m_s=1$ and $K_s=k_s$ is simply the frequency of state $s$. 

For the model based on the frequency partition $\mathcal{K}$, where subset $Q_k=\{s:~k_{s}=k\}$ for $k=1,2,\ldots$, we have $K_k=km_k$ and 
\begin{equation}
\log P\{\hat s|\mathcal{K}\} = \sum_k \log\frac{\Gamma(km_k+a)}{\Gamma(a)m_k^{km_k}}-\log\frac{\Gamma(M+aN_k)}{\Gamma(aN_k)}
\end{equation}
where $N_k=|\mathcal{K}|$ is the number of different values of $k_s$ that appear in the sample.
%

Na\"ively one would expect that $P\{\hat s|\mathcal{K}\}>P\{\hat s|\mathcal{S}\}$, i.e. that the $\mathcal{K}$ partition should always be preferred to the atomic partition $\mathcal{S}$.  \ref{proof_KS} proofs that this is indeed the case for $a=1$ and for $a\to 0$. But it also exhibit counterexamples where this is not so, in the limit of large $a$. These however correspond to rather atypical samples and no counterexample to the rule $P\{\hat s|\mathcal{K}\}>P\{\hat s|\mathcal{S}\}$ has been found in the data we have analysed. This strongly suggests that, in practical terms, the $\mathcal{K}$ partition should always be preferred to the $\mathcal{S}$ partition.

\noindent

\section{Properties of the optimal partition $Q^*$}
\label{Qalgorithm}

Finding the optimal partition $\mQ^*$ for a given sample $\hat s$ is a non-trivial task. It is reasonable to 
assume that partitions that merge states with adjacent frequencies are more likely than those that merge states with non-adjacent frequencies\footnote{If $k_{s_1}>k_{s_2}>k_{s_3}$ then a partition $\mQ$ where $s_1,s_3\in Q_{q_1}$ and $s_2\in Q_{q_2}$ will be dominated by partitions where either all three states are in different sets, or $s_1,s_2\in Q_{q_1'}'$ and $s_3\in Q_{q_2'}'$, or  $s_1\in Q_{q_1'}'$ and $s_2, s_3\in Q_{q_2'}'$, or they are all in the same set.}. Therefore, it is enough to consider partitions where all states $s\in Q_q$ have frequency $k_s$ which is larger than that of all states $s'\in Q_{q'}$ with $q>q'$. This leads us to the  following heuristics to derive the optimal partition $\mQ^*$ of a finite sample: 
\begin{enumerate}
\item Starting from $\mathcal{Q}=\mathcal{K}$:
\item For every $q=1,\ldots,N_{\mQ}-1$, define a new partition $\mathcal{Q}^{(q)}$ by merging the subsets $Q_q$ and $Q_{q+1}$ of the current partition $\mQ$ and compute the change in the log-likelihood. 
\item If the largest increase in the log likelihood over all possible values of $q$ is positive, then merge the corresponding subsets, update the partition $\mathcal{Q}$ accordingly and repeat the previous step.
\item If the largest increase in the log likelihood over all possible values of $q$ is negative, then return $\mQ^*$ as the optimal partition. 
\end{enumerate}

In order to explore the properties of $\mathcal{Q^*}$ we study ensembles where the states $s$ are drawn from power law distributions $P(s)\sim s^{-\alpha}$. This choice serves for generating data with a broad distribution of frequencies, such as those that are often observed in empirical studies. Varying $\alpha$ allows us to probe the merging algorithm proposed over a broad range of underlying distributions. 

\begin{figure}
\centering
\includegraphics[width=1\textwidth]{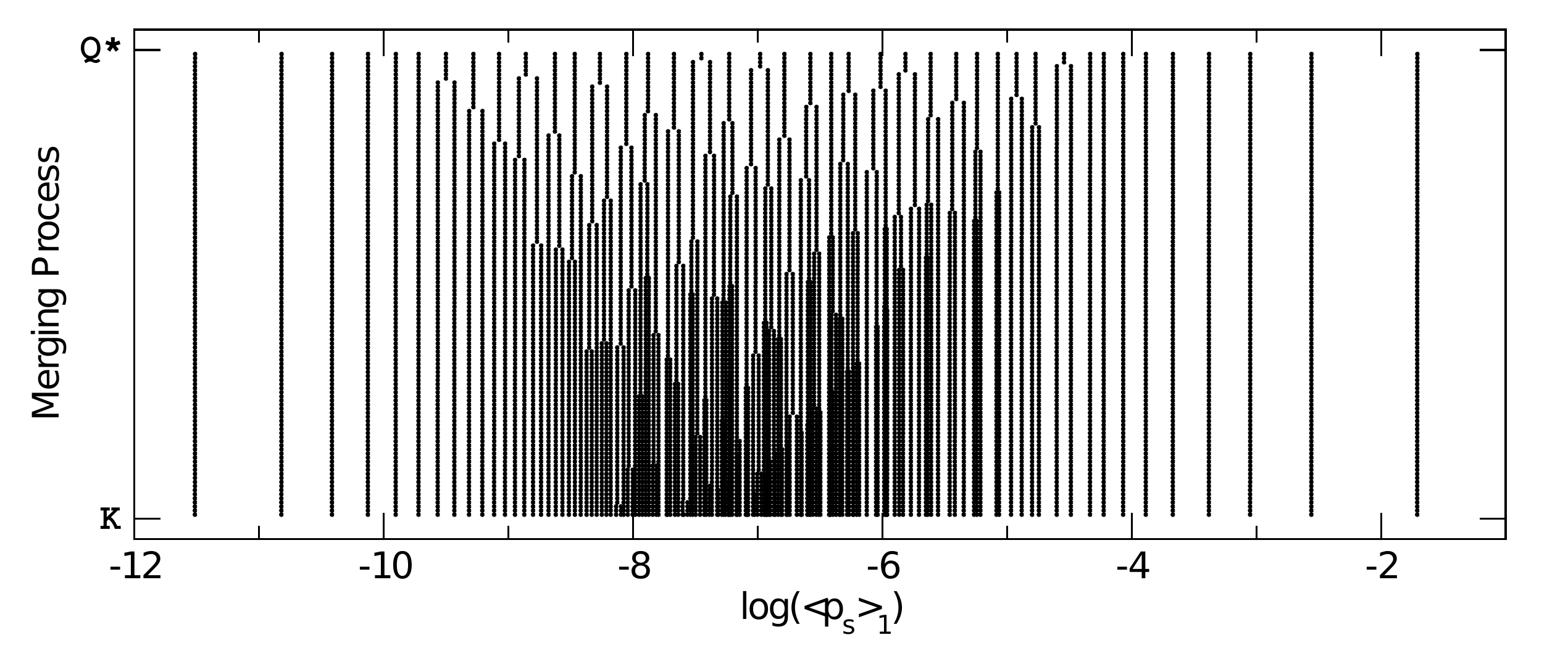}
\caption{Illustration of the Merging Process. $M=10^5$ data points were drawn from a distribution $P(s)\sim s^{-\alpha}$, with $\alpha=1.2$. The x-axis shows the estimated probability (\ref{ps_post})
for states in each subset $Q\in\mQ$. The y-axis stands for the different partitions $\mQ$ in the merging process from $\mK$ to $\mQ^*$.}
\label{barcode}
\end{figure}

Figure \ref{barcode} gives a pictorial representation of the merging process during a typical run. Interestingly, visual inspection suggests that the frequencies of the optimal model $\mQ^*$ are evenly spaced in a logarithmic scale.

One can think as well of variations to the algorithm such as selecting a favourable move at random in step (ii) instead of choosing the one that maximizes the likelihood, or merging triplets of subsets instead of pairs. We have seen that the overlaps in the final representations obtained using these variations in the algorithm are always larger than $90\%$. Moreover we see that for large samples the probability of finding a representation with $H[Q]$ greater than $H[Q^*]$ goes to zero, meaning that the later yields a more relevant description of the data. This issue is discussed in \ref{algorithms}.

\subsection{Distance between $\mathcal{Q^*}$ and $\mathcal{K}$ and scaling with the sample size}

\begin{figure}
\centering
\includegraphics[width=1\textwidth]{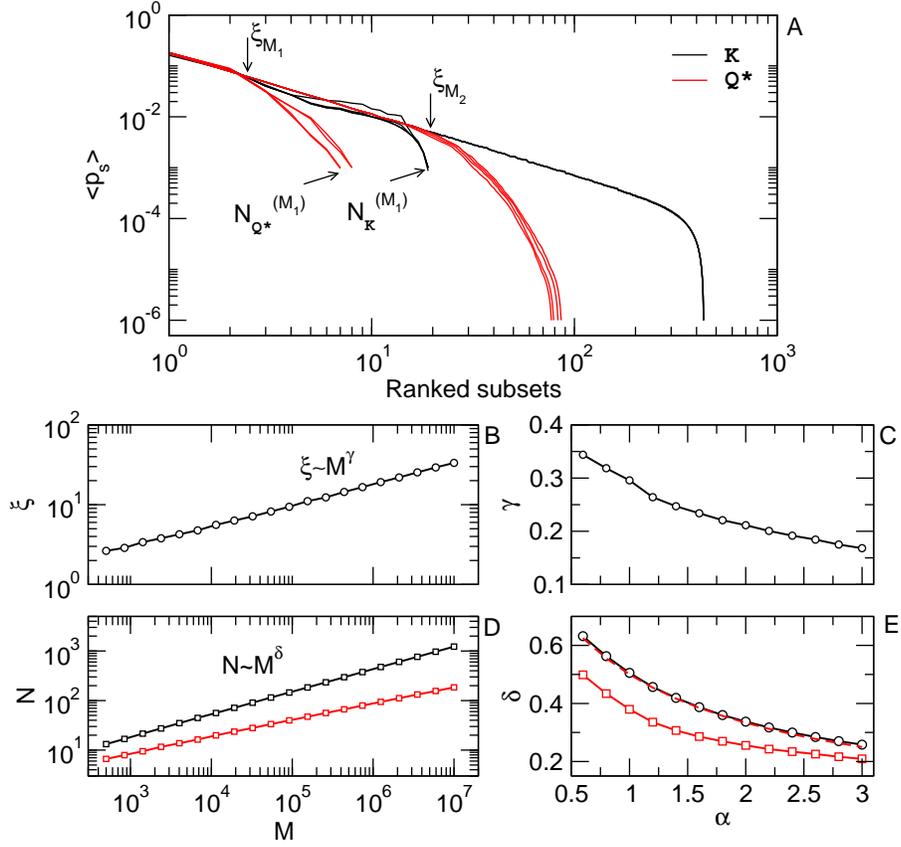}
\caption{Scaling of the optimal partition with the sample size $M$. \textbf{Panel A} shows the estimated parameters $<p_s>$ for each subset in partitions $\mathcal{K}$  and $\mathcal{Q^*}$. The data was drawn from a power law distribution $P(s)\sim s^{-\alpha}$, with $\alpha=1.2$. For both partitions we show the estimated parameters for a sample of size $M_1=10^3$ and $M_2=10^6$. $\xi$ denotes the number of parameters which are identical under both models  $\mathcal{K}$  and $\mathcal{Q^*}$. $N_{\mathcal{K}}$ and $N_{\mathcal{Q^*}}$ are the number of parameters (subsets) in each model. The different overlapping red (black) curves correspond to estimations using different values for the prior parameter $a$, ranging from 0.01 to 10. Panels B-E show analysis using $a=1$. \textbf{Panel B} shows the scaling of $\xi$ with the sample size, for $\alpha=1.2$, which follows a power law $\xi\sim M^{\gamma (\alpha)}$.  \textbf{Panel C} shows the dependence of $\gamma$ with $\alpha$. \textbf{Panel D} shows the scaling of the number of parameters in each model with the sample size, which follows a power law $N\sim M^{\delta(\alpha)}$. \textbf{Panel E} shows that the number of parameters in $\mathcal{Q^*}$ grows slower with $M$ than in $\mathcal{K}$ for a wide range of systems ($\alpha$).}
\label{N_M_scaling}
\end{figure}

Figure \ref{N_M_scaling} shows the difference between $\mathcal{Q^*}$ and $\mathcal{K}$ as a function of the sample size $M$. Panel A shows the estimated parameters $\left\langle p_s\right\rangle$ (Eq. \ref{ps_post}) for both models and two sample sizes $M_1=10^3$ and $M_2=10^6$. The states with higher frequency $k_s$ are not merged, so the partitions $\mS$, $\mK$ and $\mQ^*$ overlap on the left tail of the curve on a number $\xi$ of identical subsets of states. We estimated the $\mathcal{Q^*}$ partition and the parameters $p_s$ using priors with $a$ ranging from 0.01 to 10. The different overlapping curves in panel A stand for the different values of $a$. Clearly neither the number of subsets in $\mathcal{Q^*}$ ($N_{\mQ^*}$) nor the estimated parameters $\langle p_s\rangle$ vary strongly with $a$. In the following analysis we set $a=1$.
Panel B shows that the overlap $\xi$ between the two partitions scales with $M$ with a non-trivial exponent ($\gamma$) which depends on the underlying distribution parametrized by $\alpha$ (panel C). 
The number of parameters ($N$) in each partition gives a measure of the overfitting done in $\mathcal{K}$ with respect to $\mathcal{Q^*}$. Panel D shows that $N\sim M^\delta$ has a power law dependence on $M$ with an exponent $\delta$ that depends on $\alpha$\footnote{For the $\mathcal{K}$ partition it is possible to show that $\delta=1/(1+\alpha)$. The argument relies on the fact that the frequency of state $s$ approximates the probability $k_s/M\simeq p_s\sim s^{-\alpha}$ as long as $k_s\gg 1$ is large enough. 
We note that $m_k\simeq ds/dk$ is the number of states $s$ in an interval of frequency $dk=1$, hence $m_k\sim s^{\alpha+1}/M\sim k^{-1/\alpha-1}M^{1/\alpha}$. The number $N_k$ of states corresponds to the value of $k$ such that $m_k$ becomes of order one. Therefore $N_k\sim M^{1/(\alpha+1)}$. Interestingly, we also find that $\gamma=\delta/2$ for the $\mathcal{K}$ partition, to numerical precision. These relations do not hold for the $\mathcal{Q}$ partition.} 
(panel E). The exponent $\delta$ for the $\mQ^*$ partition is smaller than that of the $\mK$ partition implying that the difference between $N_{\mathcal{Q}^*}$ and $N_{\mathcal{K}}$ increases with $M$. 

\begin{figure}
\centering
\includegraphics[width=0.8\textwidth]{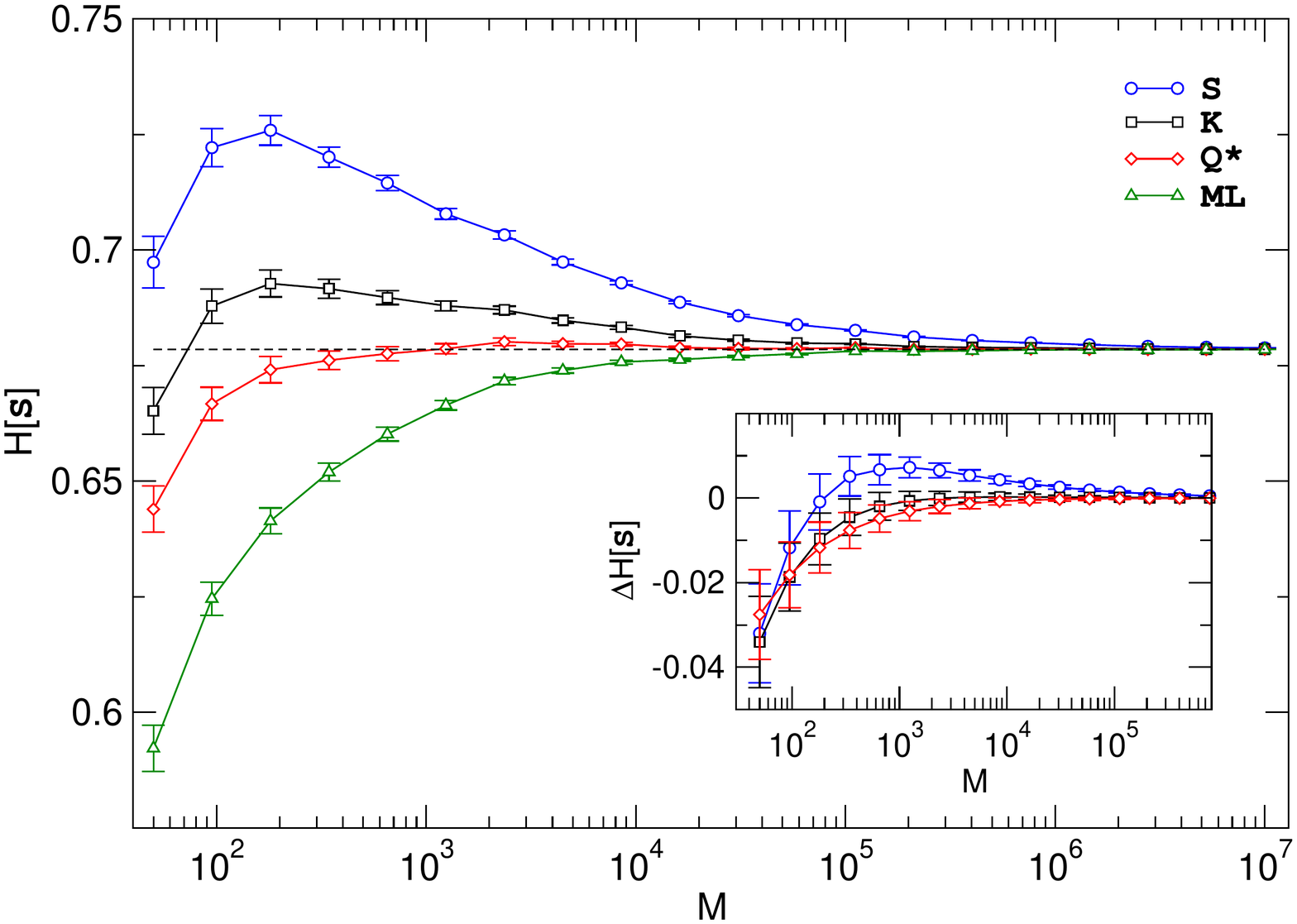}
\caption{Convergence of the estimated entropy under different models. Labels $\mathcal{S}$, $\mathcal{K}$, $\mathcal{Q^*}$  stand for the bayesian estimates of the entropy  (\ref{Hs}) in the respective model using a flat prior ($a=1$).  $ML$ stand for the maximum likelihood estimate (\ref{eqHs}). The dashed line is the true entropy of the underlying distribution $P(s)\sim s^{-3}$. Error bars denote standard errors over 1000 samples of each size $M$. The inset shows the difference between the bayesian estimates based on the posterior distribution (\ref{Hs}) and the likelihood of the model (\ref{likelihood}) $\Delta H[\mS]=\frac{1}{H^*}\left[ \left<H[\mS]|\mathcal{Q}\right>_1- (-\frac{1}{M}\log(P\{\hat s|\mathcal{Q}\}))\right] $ where $H^*$ is the true entropy of the distribution.}
\label{Hs_convergence}
\end{figure}

An interesting observation is that the optimal partition $\mQ^*$ provides an estimate of the entropy of the underlying distribution that converges faster than that based on the $\mS$ partition. The slow convergence of the entropy based on the $\mS$ partition and its strong dependence on the prior where noticed in Ref. \cite{nemenman01}, that also proposed a remedy based on treating $a$ as a hyper-parameter in Bayesian inference. Figure \ref{Hs_convergence} shows that the estimate based on the optimal partition $\mQ^*$ converges faster than finer representations, and that Bayesian inference and model selection are enough to have a reliable estimate of the entropy. This also suggests that the information kept in the coarser representation is truly relevant for characterising the sample, while the discarded information is noise associated with the finite number of data points.

\section{Criticality of maximally informative partitions}

Having provided a measure for the relevance of a given sample, allows one to characterise the typical properties of most relevant samples, i.e. of samples that are maximally informative. This question was partly addressed in Ref. \cite{marsili13}, where an upper bound to the entropy $\hat H[K]$, for a given sample size $M$ and at a given resolution $\hat H[S]$, was derived. Interestingly, this exercise shows that the distributions that achieve the upper bound in the under-sampling regime, are power laws, i.e. $m_k\sim k^{-\mu-1}$. This suggests that ``criticality'', i.e. the observation of scale-free frequency distribution, may be a consequence of choosing the most informative variables, and need not necessarily imply underlying mechanisms of self-organisation to a critical point.

In \ref{poissonisation} we revisit the argument leading to the upper bound and also derive a lower bound for $\hat H[K]$, showing that this is also achieved when the distribution of frequencies has a power law behaviour $m_k\sim k^{-\mu-1}$. 

The observation (see Fig. \ref{barcode}) that model selection identifies partitions $\mQ^*$ with posterior probabilities $\langle p_s|\mQ^*\rangle_1$ that are evenly spaced on a logarithmic scale, suggests that the same may be true for samples of a given size $M$, with a maximal $\langle H[Q]|\mQ^*\rangle_1$ at a given resolution $\langle H[S]|\mQ^*\rangle_1$. 

Yet, in order to further corroborate this conclusion, one needs to resort to numerical simulations. To this end, we generated samples from Montecarlo simulations maximising the measures of relevance proposed above. 
The simulations consisted in the following steps: 

\begin{enumerate}
\item Start with an arbitrary sample defined by the frequencies $\hat k=(k_1,...,k_N)$, with $\sum_s^N k_s=M$, and $N$ the initial number of states. Without loss of generality sort the frequencies in decreasing order $k_1\ge ...\ge k_N$.
\item Consider every possible move of $n$ samples from state $i$ into state $j\ne i$ for all $i\in [1,N]$,$j\in [1,N+1]$ and $n$, under the constraint that $k_i-n\ge k_{i+1}$ and $k_{j-1}\ge k_j+n$. Notice that $j=N+1$ implies defining a new state with frequency $n$. Conversely if $k_i=1$ this state will disappear after moving the one sample to state $j$.
\item Choose the move which maximizes the Lagrange function $\mathcal{L}=H[Q]+\mu H[S]$, with $Q=K$ and $Q=Q^*$ independently, for a fixed value of $\mu$.
\item Repeat (ii)-(iii) until there is no favourable move.
\end{enumerate}

\begin{figure}
\centering
\includegraphics[width=0.9\textwidth]{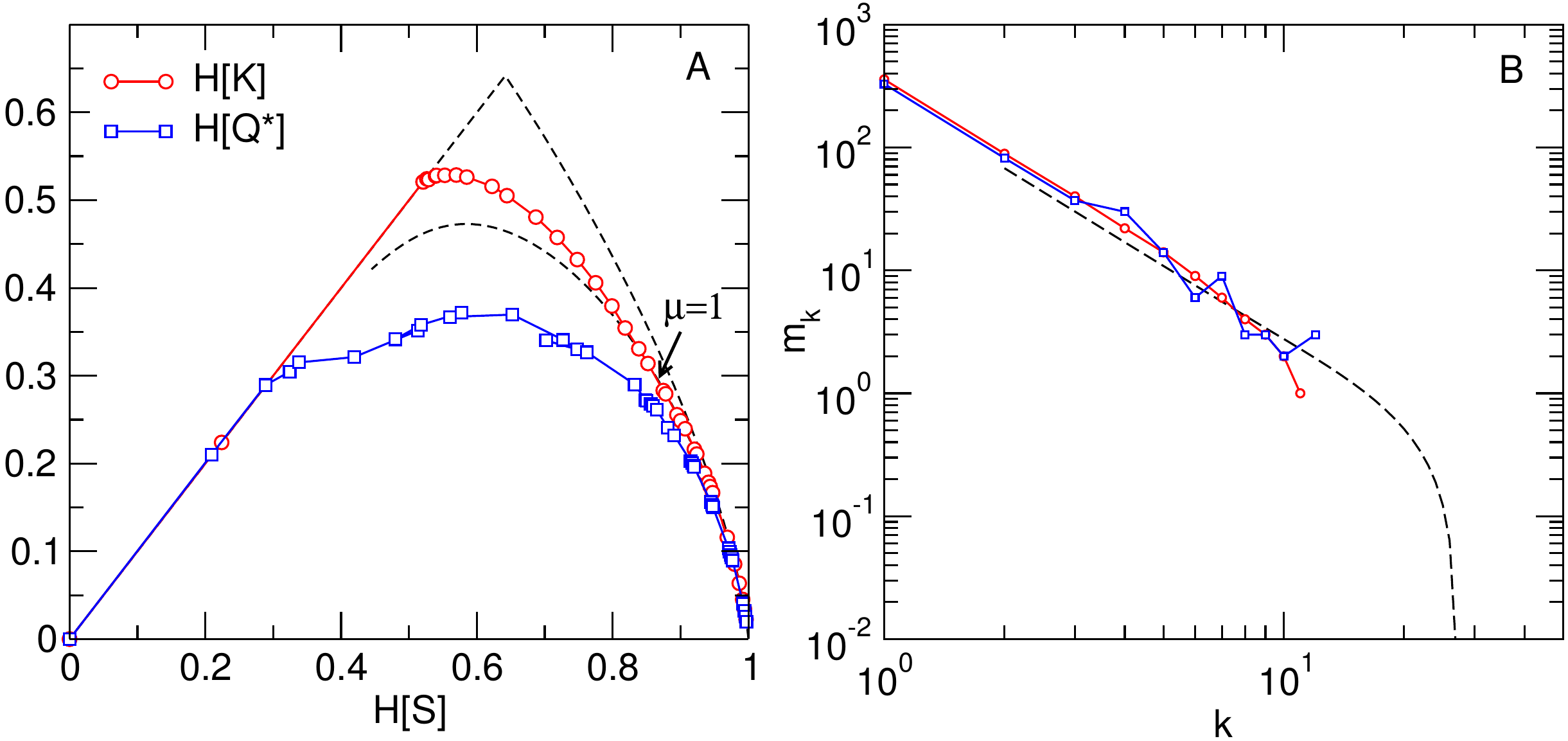}
\caption{Maximally relevant samples. Samples of size M=1000 were generated via Monte Carlo simulations maximizing $H[K]+\mu H[S]$ and $H[Q^*]+\mu H[S]$ (see main text), for fixed values of $\mu \in [-1,5]$ which determine the resolution $H[S]$ of the solutions. All entropies are normalised by $\log(M)$. Panel A shows the maximized $H[K]$(red) and $H[Q^*]$(blue) together with theoretical bounds for $H[K]$ (dashed). The solutions expected are of the form $m_k \sim k^{-\mu-1}$. Panel B shows the solutions obtained for $\mu=1$, which are power laws of exponent 2 (Zipf's law). The dashed line is an approximate solution to the theoretical lower bound for $H[K]$ where $m_k$ is assumed to be a poisson variable (see \ref{poissonisation}).} 
\label{MC_samples}
\end{figure}

Keeping $\mu$ fixed in step (iii) and allowing H[S] to fluctuate accordingly during the simulation favoured the ergodicity of the dynamics with respect to fixing H[S] and maximizing H[Q]. We repeated the simulations for different values of $\mu \in [-1,5]$. For each value of $\mu$ we repeated the simulations with different initial conditions. This was not essential for the maximization of H[K] but for the maximization $H[Q^*]$ the process converged to local maxima strongly dependent on the initial conditions. We therefore varied the initial number of states $N$ from 25 to 950, for a sample of size $M=1000$, and performed 20 independent realizations for each initial resolution. The absolute maximum of $H[Q^*]+\mu H[S]$ across realizations was kept for each $\mu$. Panel A in figure \ref{MC_samples} shows the results for both relevance measures $H[K]$ and $H[Q*]$.  Values of $\mu<-1$ yield the trivial result of $H[Q]=0$ and $H[S]=0$ which corresponds to the solution $k_1=M$, $k_{j>1}=0$. $\mu=-1$ yields solutions with $H[Q]=H[S]$ in the well sampled regime (left part of the diagram). In the case of $Q=K$, the solutions are of the form $m_k\sim k^{-\mu-1}$ (see \ref{poissonisation}). Panel B shows the solutions obtained for $\mu=1$ which match the expected Zipf law. The dashed curves in panel A refer to theoretical upper and lower bounds for the value of $H[K]$ (\ref{poissonisation}). 

\section{Application to real data}

In this section we compare the models based on the $\mathcal{K}$ and $\mathcal{Q^*}$ partitions in two applications to real data. The $\mathcal{K}$ partition is derived directly and exactly from the data whereas the $\mathcal{Q^*}$ partition requires a calculation that may be heavy and approximate. The scope of this section is to show that in practical cases, the $\mathcal{K}$ partition is a very good approximation to the optimal one $\mathcal{Q^*}$. Intuitively, the reason why this is so relies on the fact that informative samples (those with a large $H[Q^*]$ or $H[K]$) have broad frequency distributions, and as we have seen, the $\mQ^*$ and $\mK$ partitions have a sizeable overlap in these cases.

\begin{figure}
\centering
\includegraphics[width=1\textwidth]{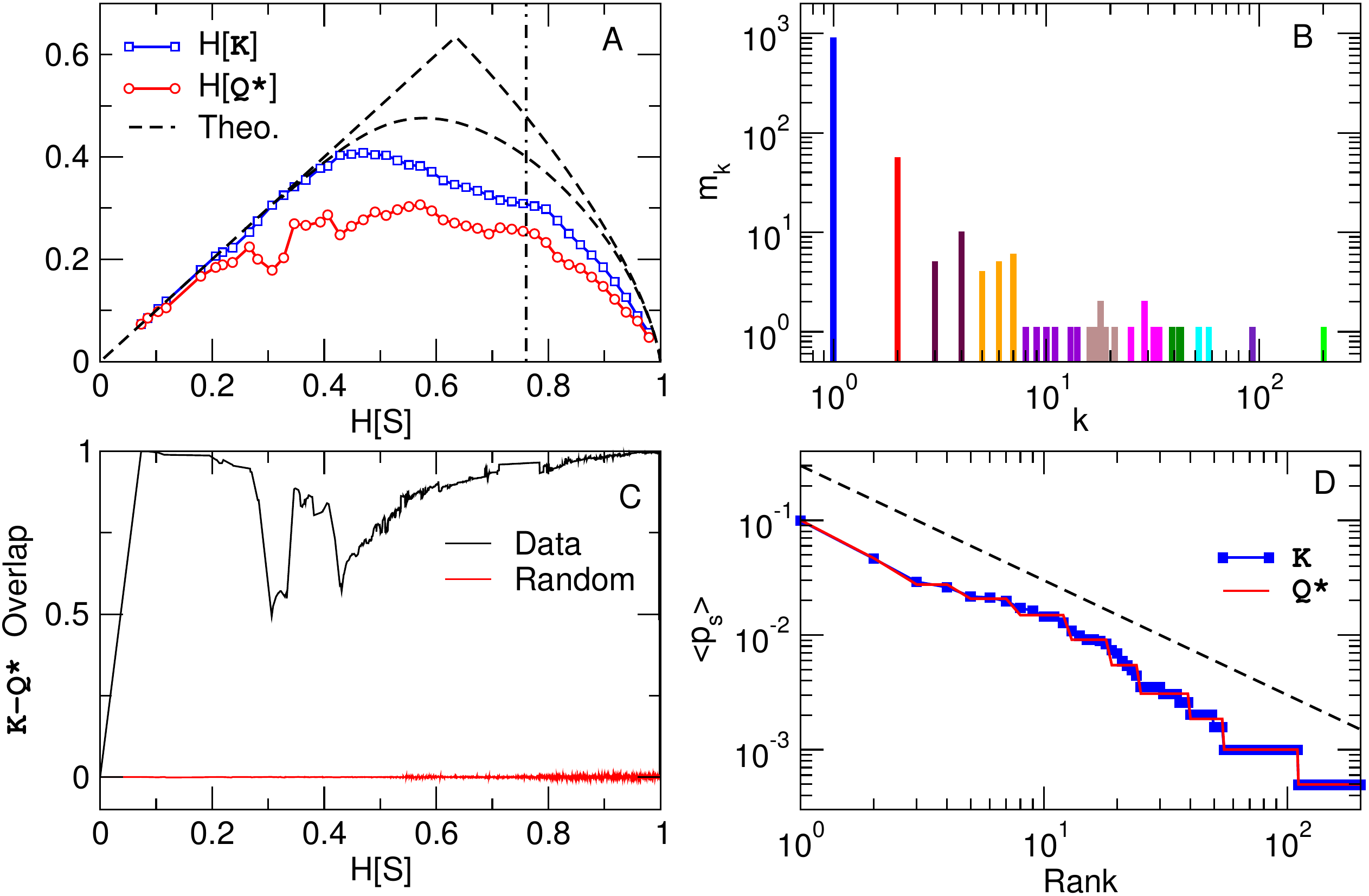}
\caption{$\mK$ and $\mQ^*$ representations of a financial market data set of stock returns. Stocks where clustered at different resolutions ($H[S]$) by means of the algorithm developed in Ref. \cite{marsili02b}. The relevance of each level of description is quantified by $H[K]$ and $H[Q^*]$ in Panel A (all entropies are normalized by $\log(M)$). Panel C shows the overlap between partitions $\mK$ and $\mQ^*$ at each resolution. The overlap index is the Adjusted Rand Index (ARI, Ref. \cite{hubert85}) which yields 1 for identical partitions and 0 when the overlap is expected by chance. The partition labels where shuffled before computing the ARI (red curve) to illustrate this case. Panels B and D refer to partitions $\mK$ and $\mQ^*$ of the data, at the resolution marked by the vertical dashed line in A ($H[S]\sim 0.75$). Bars in Panel B show the $\mK$ partition. The x-axis are the frequencies and the y-axis are the number of states seen with each frequency. The colours show the coarser $\mQ^*$ partition, obtained by the algorithm of section \ref{Qalgorithm}. Panel D shows the estimated parameters with the model based on $\mK$ and $\mQ^*$. The dashed line is a Zipf law for comparison.}
\label{Financial}
\end{figure}

In the first example, we analyse a financial market data set of stock returns. The data set (used previously in \cite{marsili02b,marsili02a}) span a period from 1st January 1990 to 30th April 1999 (2249 time points) and it covers the $M=2000$ most frequently traded stocks in the New York Stock Exchange in that period. Assuming that returns are gaussian with a block diagonal correlation matrix allows one to group stocks in clusters of ``sectors'', by maximum likelihood (see \cite{marsili02a,marsili02b} for details). 
The cluster label $s_i$ of each stock $i=1,\ldots,M$ identifies the $\mS$ partition in this context. As the number $N_s$ of clusters varies from $1$ to $M$, the algorithm produces partitions $\mS$ with a different resolution $H[S]$. We compare the relevance of different levels of description by computing $H[K]$ and $H[Q^*]$. Here the optimal partition $\mQ^*$ is obtained with the algorithm defined in Section \ref{Qalgorithm} starting from $\mK$. 
Panel A in figure \ref{Financial} shows both measures of relevance as a function of the resolution $H[S]$. The dashed curves are theoretical upper and lower bounds to the estimate of the maximal value of $H[K]$, given $H[S]$ and $M$ (see \ref{poissonisation}). Panel B illustrates the relation between the $\mK$ and $\mQ^*$ representations, at the resolution marked by the vertical dashed line in panel A. Bars in panel B denote the $\mK$ partition, whereas the colours indicate which frequencies were merged together to form the coarser optimal model $\mQ^*$.
Panels C and D provide a closer look at the distance between partitions $\mK$ and $\mQ^*$. Panel C shows the overlap between the $\mK$ and $\mQ^*$ partitions at each resolution. The overlap was computed by the Adjusted Rand Index (\cite{hubert85}), which is bounded above by 1 and yields 0 when the overlap matches the one expected by chance. For illustrative purposes we show the overlap between shuffled versions of the partitions (red curve), which indeed yield constant zero for all resolutions. We point out that in the strongly under-sampled domain, both models are practically the same.
Panel D shows the estimated parameters (Eq. \ref{ps_post}) in the $\mK$ and $\mQ^*$ models, at the resolution marked by the vertical line in A. The dashed line is a Zipf law for comparison. 


\begin{figure}
\centering
\includegraphics[width=0.9\textwidth]{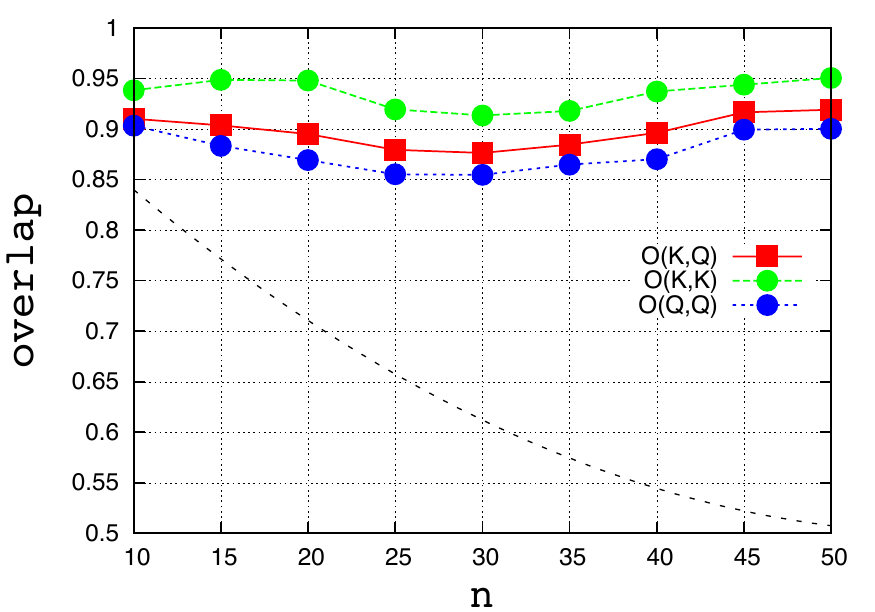}
\caption{Self-overlap $O_{\mK}$ and $O_{\mQ^*}$ (filled circles) of the subset of site $I^*_{\mK}$ and $I^*_{\mQ^*}$ produced by the optimisation of $H[K]$ and $H[Q^*]$, respectively, on the MSA for the voltage sensor protein domain. Self-overlaps are computed as the probability that a randomly chosen site is either selected or not in two different runs of the algorithm of Ref. \cite{grigolon15} and are shown, in the plot, as a function of the number $n$ of sites in $I^*$. The overlap $O(K,Q)$ between the subsets $I^*_{\mK}$ and $I^*_{\mQ^*}$ is also computed in the same way, and it is shown with filled squares. The dashed line corresponds to the overlap between subsets of $n$ sites chosen randomly among the $L=114$ possible sites.} 
\label{Protein}
\end{figure}

As a second example, following Ref. \cite{marsili13}, we consider the problem of identifying relevant positions in the sequence of amino acid that correspond to a particular protein domain. In brief, the data consists of Multiple Sequence Alignments (MSA) of $M$ sequences of the same protein domain, across different species. We refer to \cite{grigolon15} for a detailed description, for our purposes here, suffice it to say that a protein domain can be identified by a sequence $\vec a=(a_1,\ldots,a_L)$ of $L$ amino acids, each being of one of 21 possible types (e.g. $a_i=V$ for valine, $a_i=A$ for alanine, etc) and that an MSA is a collection of $M$ such sequences across different organisms or species. The key point is that, while the whole sequence is subject to a random process of mutations, there are features which need to be conserved in order to perform the function the protein is supposed to do. In order to understand which positions along the sequence are relevant for the biological function, we observe that each subset $I\subseteq \{1,2,\ldots,L\}$ of the positions identifies a partition $\mS_i$ of the MSA data, whose elements $s=(a_{i},i\in I)$ are the subsequences of the domain on the positions $i\in I$. 
From this we can define the $\mK$ and the optimal $\mQ^*$ partitions, and compute both the resolution  $H_I[S]$ and the relevance $H_I[Q^*]$ or $H_I[K]$ corresponding to this subset of positions. 
This makes it possible to look for the most relevant subset of positions $I^*$, as the one that maximises $H_I[Q^*]$. 
This program is carried out in Ref. \cite{grigolon15} to which we refer the interested reader. Here we confine the discussion to the comparison of the $\mK$ and $\mQ^*$ partitions. In brief, the maximisation (of either $\mK$ or $\mQ^*$) is done using a Montecarlo algorithm for subsequences of a fixed number $n$ of amino acids. We applied the algorithm to the Voltage Sensor Domain of ion channels (Pfam code PF000520). The data was the same as that used in Klein {\em et al.}  \cite{MSApaper}. In brief, the algorithm of Ref. \cite{grigolon15} produces a distribution over the subsets $I^*$ of $n$ relevant sites that allows one to compute the probability that a site is either selected or not in two different realisations of $I^*$. Fig. \ref{Protein} shows that optimising the relevance of the $\mK$ or the $\mQ$ partitions provides a sharp separation between relevant and irrelevant positions, which is sharper for the $\mK$ partition. In addition, the selected subsets of sites $I^*_{\mK}$ and $I^*_{\mQ^*}$ under the optimisation of $H[K]$ or $H[Q^*]$ have a large relative overlap: in 90\% of the cases, the two optimisation schemes yield the same prediction on whether a site that is relevant or not (see Fig. \ref{Protein} and the caption for details). 

These two examples suggest that, in practical applications, $H[K]$ can be used as a proxy for $H[Q^*]$ in measuring the relevance. This is particularly useful to avoid the optimisation leading from $\mK$ to $\mQ^*$ and speed up numerical calculations.

\section{Conclusion}

The Big Data revolution has made available data of unprecedented detail on the working of complex systems, such as cells, networks of neurons and the brain, ecologies, social networks, economies and financial markets. This, in particular, indicates that quantitative approaches typical of hard sciences can be extended to life sciences as well. Yet, the fact that such phenomena are not constrained by well known laws, as in physics, makes inference of behaviour a daunting task. Indeed, one is rarely in the circumstance where behaviour depends on only few variables that can be controlled. In such cases, the resolution of high dimensional data, is not given by the number of variables that one can measure simultaneously, but rather is limited by statistical errors induced by finite sample size. Dimensionality reduction schemes have to be invoked to adjust the resolution so that reliable statistical information can be extracted from the data. This inevitably introduces a tradeoff between relevance and resolution, which is addressed in this paper.

The main contribution of this paper is to make this tradeoff explicit and quantitative in information theoretic terms, on the basis of a Bayesian model selection approach. We focus on the limiting case where the system under study is severely under-sampled and no other information apart the frequency of observations is available. There, models are in one to one correspondence with partition of the set of observed states. So while resolution is a measure of the number of different states, relevance can be defined in terms of the number of different elements in the partition, i.e. the number of different states that the data allows one to distinguish. We find that, as resolution increases from the coarser possible level, relevance increases up to a maximum, beyond which it starts decreasing. In the extreme limit where each observation is seen only once, relevance vanishes, signalling that data contains no relevant information on the system. 

The resolution (i.e. the number of sets in the partition) also provides a natural cutoff in the number of parameters that the data allows us to infer, beyond which inference would result in overfitting. The number of parameters (and of partitions) increases with the sample size $M$. Loosely speaking, as $M$ increases, the model passes through a sequence of symmetry breaking transitions where more and more distinctions between states can be made. This process, indeed, bears well known formal analogies with the symmetry breaking process in physical systems when the temperature (here proportional to $1/M$) decreases.

There are several interesting directions for further research along these lines. One is to extend the approach in Section \ref{sec:unsampled} to explore sampling processes \cite{goodturing} that are consistent with the Bayesian model selection scheme. The second is to exploit these results for inference of graphical models in cases where states can be considered as a configuration $s=(\sigma_1,\ldots,\sigma_n)$ of an extended system. There, a well established technique is Boltzmann learning (see e.g. \cite{schneidman06}) which, given a set of relevant observables, invokes maximum entropy principle and predicts a distribution $P(\sigma_1,\ldots,\sigma_n)$. The set of relevant observables determines the model. Yet, no general criterium exists for dictating what relevant observables should be and it seems natural to invoke model selection schemes to address the issue. 

Finally, the present approach also suggests a new perspective on the widespread occurrence of criticality. It suggests that the occurrence of broad frequency distributions is a consequence of sampling relevant variables in the under sampling regime. In this spirit the interesting question is not whether or why "biological systems are poised at criticality" \cite{MoraBialek} but rather how to use the "apparent criticality" of frequency distributions to select relevant variables.

\section{Acknowledgements} 

We thank W. Bialek, A. Celani, P. Latham, Y. Roudi and M. Vergassola for interesting discussions. 
This work was supported by the Marie Curie Training Network NETADIS (FP7, 290038).


\appendix 

\section{Samples that maximise $\hat H[K]$ have power law distribution}\label{poissonisation}

The problem is to find the distributions $m_k\in \mathbb{N}$ that satisfy
\begin{equation}
\label{ }
\sum_{k} km_k=M,\qquad \hat H[s]\equiv -\sum_k \frac{km_k}{M}\log \frac{k}{M}=H_0
\end{equation}
and maximize 
\begin{equation}
\label{ }
\hat H[K]\equiv -\sum_k \frac{km_k}{M}\log \frac{km_k}{M}.
\end{equation}
The problem is difficult because it has to be solved for integer $m_k$. In order to circumvent this problem we 
think of $m_k$ as being drawn from a distribution and maximise the expected value of $\hat H[K]$, subject to 
the constraints that the expected value of $\hat H[s]$ and $M=\sum_k k m_k$ are fixed. 
The main technical problem relies in computing the expected value of $m_k\log m_k$. On one side, one can 
observe that 
\[
E\left[  m_k\log m_k\right]\ge n_k\log n_k ,\qquad n_k=E[m_k].
\]
This makes it possible to derive an upper bound on the maximal value of $\hat H[K]$. Indeed, one particular 
distribution of $m_k$ is one where $m_k=n_k$ for all $k$, with integer $n_k$.  The maximisation over these 
distributions coincides with the original problem. Maximising 
\[
\hat H_{\rm ann}[K]=-\sum_k\frac{kn_k}{M}\log\frac{kn_k}{M}
\]
over all real $n_k\ge 0$ with $\sum_k km_k=M$ and $\sum_k{kn_k}\log({k}/{M})=-{M}H_0$, clearly produces 
an upper bound to the true solution. This upper bound, as discussed in \cite{marsili13} predicts power law distributions 
$m_k\sim k^{-\mu-1}$ with $\mu\ge 1$.

In order to derive a lower bound, we confine ourselves to a specific class of distributions.
More precisely, we take $m_k$ as Poisson variables with mean $n_k$ and solve the problem of finding $n_k$ such that the average of $\hat H[K]$ is maximised under the same constraints as above.
Notice that this is akin to studying the problem in the analog Gran Canonical Ensemble where $M$ is a allowed to fluctuate.
What we need to check a posteriori is that the fluctuations of $M$ are small compared to the mean.

The only nontrivial part of the calculation has to do with computing the expected value of $m_k\log m_k$, for which we use the formula
\begin{equation}
\label{ }
\log z=\int_0^\infty\frac{du}{u}\left(e^{-u}-e^{-zu}\right)
\end{equation}
so that, for a Poisson variable $m$ with mean $n$, we find

\begin{eqnarray}
E[m\log m] & = & n\int_0^{\infty} \frac{du}{u}\left(1-e^{-n(1-e^{-u})}\right) \\
 & = & n\int_0^1dz\frac{e^{-nz}-1}{\log(1-z)}\\
 & = & \int_0^ndt\frac{e^{-t}-1}{\log(1-t/n)} 
\end{eqnarray}
The first expression can be used to check that 
\[
E[m\log m] \simeq a n^2+O(n^3), \qquad a=-\int_0^1dz\frac{z}{\log(1-z)}
\]
for $n\ll1$, whereas the last shows that $E[m\log m]\simeq n\log n$ for $n\gg 1$.

Writing $E[F]=\mathcal{F}$ we find
\begin{equation}
\label{ }
\mathcal{F}=-\sum_k \frac{k n_k}{M}\left[\mathcal{L}(n_k)+(\mu+1)\log\frac{k}{M}-\lambda\right]-\mu H_0-\lambda M
\end{equation}
where
\begin{equation}
\label{ }
\mathcal{L}(n)=\int_0^1dz\frac{e^{-nz}-1}{\log(1-z)}
\end{equation}
Notice that the only problematic thing here is that we are not taking into account that $M$ also is a random variable. Operationally, one could think at taking an ensemble of systems, $m_k^{(a)}$ all strictly satisfying the constraint. Then we define the ensemble average $n_k$ of the $m_k$'s and pretend that its distribution be Poisson, which seems reasonable.

The extrema of $\mathcal{F}$ can now be computed: $n_k$ will satisfy
\begin{equation}
\label{FOC}
n_k\mathcal{L}'(n_k)=\lambda-(\mu+1)\log\frac{k}{M}-\mathcal{L}(n_k)
\end{equation}
that can be solved numerically foe each $k$.

Notice that $P\{m_k>0\}=1-e^{-n_k}$, therefore the expected number $N$ of states $s$ visited is
\begin{equation}
\label{ }
N=\sum_k (1-e^{-n_k})
\end{equation}

In order to compute $k_{\max}$ notice that 
\begin{equation}
\label{ }
P\{k_{\max}< q\}=\prod_{k=q}^\infty P\{m_k=0\}=e^{-\sum_{k>q} n_k}
\end{equation}
Finally, the variance of $M$ is given by
\begin{equation}
\label{ }
V(M)=\sum_k k^2  V(m_k)=\sum_k k^2  n_k
\end{equation}
and the validity of the method relies on the fact that 
\begin{equation}
\label{ }
\lim_{M\to\infty} \frac{V(M)}{M^2}=0
\end{equation}

A rather crude approximation of the solution is possible if we take
\begin{equation}
\label{ }
n_k\mathcal{L}'(n_k)+\mathcal{L}(n_k)\approx\log(1+n_k/n_c)=-(\mu+1)\log\frac{k}{k_c},\qquad k_c=Me^{-\lambda/(\mu+1)}
\end{equation}
for $k<k_c$ and $n_k=0$ for $k\ge k_c$. With $n_c=0.38$ the approximation is valid to less than 1\% for $n>10$ but it underestimates by 80\% the true vale at small $n$ (a larger value of $n_0$ would give a best fit to the small $n$ region).

Within this approximation
\begin{equation}
\label{ }
n_k=n_c\left(\frac{k}{k_c}\right)^{-\mu-1}
\end{equation}
and it is consistent to take
\begin{equation}
\label{ }
\mathcal{L}(n)\approx\left(1+\frac{n_c}{n}\right)\log\left(1+\frac{n}{n_c}\right)-1
\end{equation}
Therefore
\begin{eqnarray}
\hat H[s] & \approx & \sum_{k=1}^{k_c}\frac{k n_k}{M}\log\frac{k}{M} \\
\hat H[K] & \approx & \hat H[s]+\sum_{k=1}^{k_c}\frac{k n_k}{M}\left[\left(1+\frac{n_c}{n_k}\right)\log\left(1+\frac{n_k}{n_c}\right)-1\right]
\end{eqnarray}


\section{Properties of $\mathcal{L}(n)$}

For small $n$:
\begin{eqnarray}
\label{ }
\mathcal{L}(n)=\int_0^1dz\frac{e^{-nz}-1}{\log(1-z)} \\
\simeq \log (2) n-\frac{1}{2}\log\left(\frac{4}{3}\right) n^2+
\frac{1}{6}\log\left(\frac{32}{27}\right) n^3-\frac{1}{24}\log\left(\frac{4096}{3645}\right) n^4+O(n^5) \nonumber
\end{eqnarray}
We can write

\begin{equation}
\label{ }
\mathcal{L}(n)=\int_0^\infty\!dx e^{-x}{\rm li}(1-x/n)=\log n +\int_0^n\frac{dz}{z}\left[e^{-z}-e^{-z/n-n(1-e^{-z/n})}\right]
\end{equation}
where ${\rm li}(x)=E_1(\log x)$ is the logarithmic integral function.

\section{Comparison between the $\mathcal{K}$ and the $\mathcal{S}$ 
partitions}\label{proof_KS}

The partition $\mathcal{K}$ is clearly preferable to $\mathcal{S}$ in the limit $a\to 0$, as the likelihood ratio behaves as $a^{N_s-N_k}$. We first argue that this is also the case for $a=1$ (uniform prior) and then we analyse the opposite limit $a\to\infty$. 

Consider the $\mathcal{K}$ partition of size $N$ for $a=1$. Suppose that there are $m$ states that occur with frequency $k$, being therefore in the same subset in $\mathcal{K}$. Consider now a new partition $\mathcal{Q}$ in which we have atomised one of the $m$ states to a new subset of size 1. We will show that the likelihood of the $\mathcal{Q}$ model is smaller than the one of $\mathcal{K}$

\begin{equation} \label{KvsS}
\frac{P\{\hat s|\mathcal{K}\}}{P\{\hat s|\mathcal{Q}\}}>1
\end{equation}
for $a=1$.

Using Eq. (\ref{likelihood}), equation (\ref{KvsS}) takes the form
\begin{eqnarray}
\frac{P\{\hat s|\mathcal{K}\}}{P\{\hat s|\mathcal{Q}\}}&=   f(k,m)g(M,N)\\
f(k,m)&= \frac{(km!)}{(k(m-1))!k!} \frac{1}{m^k} \left( 1-\frac{1}{m}\right)^{k(m-1)} \end{eqnarray}
where $g(N,M)=\frac{M+N}{N}$ is an increasing function of $M$ and it decreases with $N$. So the worst case scenario is when $M$ is small and $N$ is large. This corresponds to an original $\mathcal{K}$ partition with $N-1$ subsets of size $m_k=1$ and $k={1,2,...,N-1}$, plus the one subset of size $m$ and frequency $k=N$ from which we are atomising one state. This yields the smallest value of $M$, compatible with $k,m$ and $N$, which is 
\begin{equation}
M^* =km+\frac{N(N-1)}{2}. \label{minM}
\end{equation}
This gives $g(N,M^*)= \frac{N+1}{2}+\frac{km}{N}$. The minimal value of $g$ is now obtained for $N^*= \sqrt{2km}$, which implies that
\[
g(M,N)\ge g(M^*,N^*)=\sqrt{2km}+\frac{1}{2}.
\]
On inspection, it is easy to check that $f(k,m)\cdot g(M^*,N^*)$ is an increasing function of $m$, so it attains its minimum value at $m=2$. Therefore
\begin{equation}
\label{ }
\frac{P\{\hat s|\mathcal{K}\}}{P\{\hat s|\mathcal{Q}\}}\ge \frac{2}{\sqrt{\pi}}+\frac{1}{2\sqrt{\pi k}}> \frac{2}{\sqrt{\pi}}=1.128\ldots > 1.
\end{equation}
Notice that the worst case limit of $m=2$ is attained when the $\mathcal{Q}$ partition becomes exactly $\mathcal{S}$.

Yet, in the limit of large $a$, the ratio of the likelihood may become less than one. In order to address this issue, we shall exhibit a specific case for $a\to \infty$.

Let us split the log-likelihood ratio in three pieces:
\begin{eqnarray}
\Delta(a)=\log \frac{P\{\hat s|\mathcal{K}\}}{P\{\hat s|\mathcal{S}\}} & = & 
\sum_k \log\frac{\Gamma(km_k+m_ka)/\Gamma(m_k a)}{[\Gamma(k+a)/\Gamma(a)
m_k^{k}]^{m_k}} \\
&~&-\sum_{k}\log\frac{\Gamma(km_k+m_k a)/\Gamma(m_k a)}{\Gamma(km_k+a)/\Gamma(a)}\nonumber \\
&~&+\log\frac{\Gamma(M+aN_s)/\Gamma(aN_s)}{\Gamma(M+aN_k)/\Gamma(aN_k)} \nonumber
\end{eqnarray}
Writing $\Delta=\Delta_1+\Delta_2+\Delta_3$, that correspond to the three lines above, using Stirling's approximation, it is easy to show that
\begin{eqnarray}
\Delta_1 & \simeq & \sum_{k:m_k\ge 1}\frac{(m_k-1)k}{2a}+O(a^{-2}) \\
\Delta_2 & \simeq & -\sum_k k m_k\log m_k +\sum_{k:m_k\ge 1}\frac{m_k(m_k-1)k^2}{2a}+O(a^{-2})\\
\Delta_3 & \simeq & M\log\frac{N_s}{N_k}-(N_k^{-1}-N_s^{-1})\frac{M^2}{2a}+O(a^{-2})
\end{eqnarray}

%
%
The leading order term can be cast in the form
\begin{equation}
\label{ }
\Delta=M\left[\log N_s-\hat H[S]\right]-M\left[\log N_k-\hat H[K]\right]
\end{equation}
The first is the amount of information, in nats, that one gains from the knowledge of $p_s=k_s/M$ (over the uniform distribution on $s$) whereas the second is the amount of information one gains from the knowledge of $p_k=km_k/M$ (over the uniform distribution on $k$). It seems intuitive that the first is larger then the second. 

Yet it is easy to find counterexamples: Take a sample with $M=mk+k_0$ points, $m$ states occur $k_s=k$ times and one occurs $k_0$ times, therefore $N_s=m+1$ and $N_k=2$. Then $p_k=1/(1+x)$ and $p_{k_0}=x/(1+x)$, with $x=k_0/(mk)$ and $p_s=(k/M,\ldots,k/M,k_0/M)$. Then 
\[
\hat H[S]-\hat H[K]=\frac{1}{1+x}\log m, \qquad \Delta =\log \frac{m+1}{2}-\frac{1}{1+x}\log m
\]
Then $\Delta<0$ for
\[
k_0\le mk\frac{\log[2m/(m+1)]}{\log[(m+1)/2]}
\]
For $m=2$ this occurs for $k_0<1.419 \cdot k$, for $m=3$ $k_0<1.755 \cdot k$ and for 
$m=10$ $k_0<3.507\cdot k$. These, however seem rather pathological samples that will not typically arise in a sampling process.

\section{Variations in the algorithm for defining $\mathcal{Q}^*$}\label{algorithms}

\begin{figure}
\centering
\includegraphics[width=0.5\textwidth]{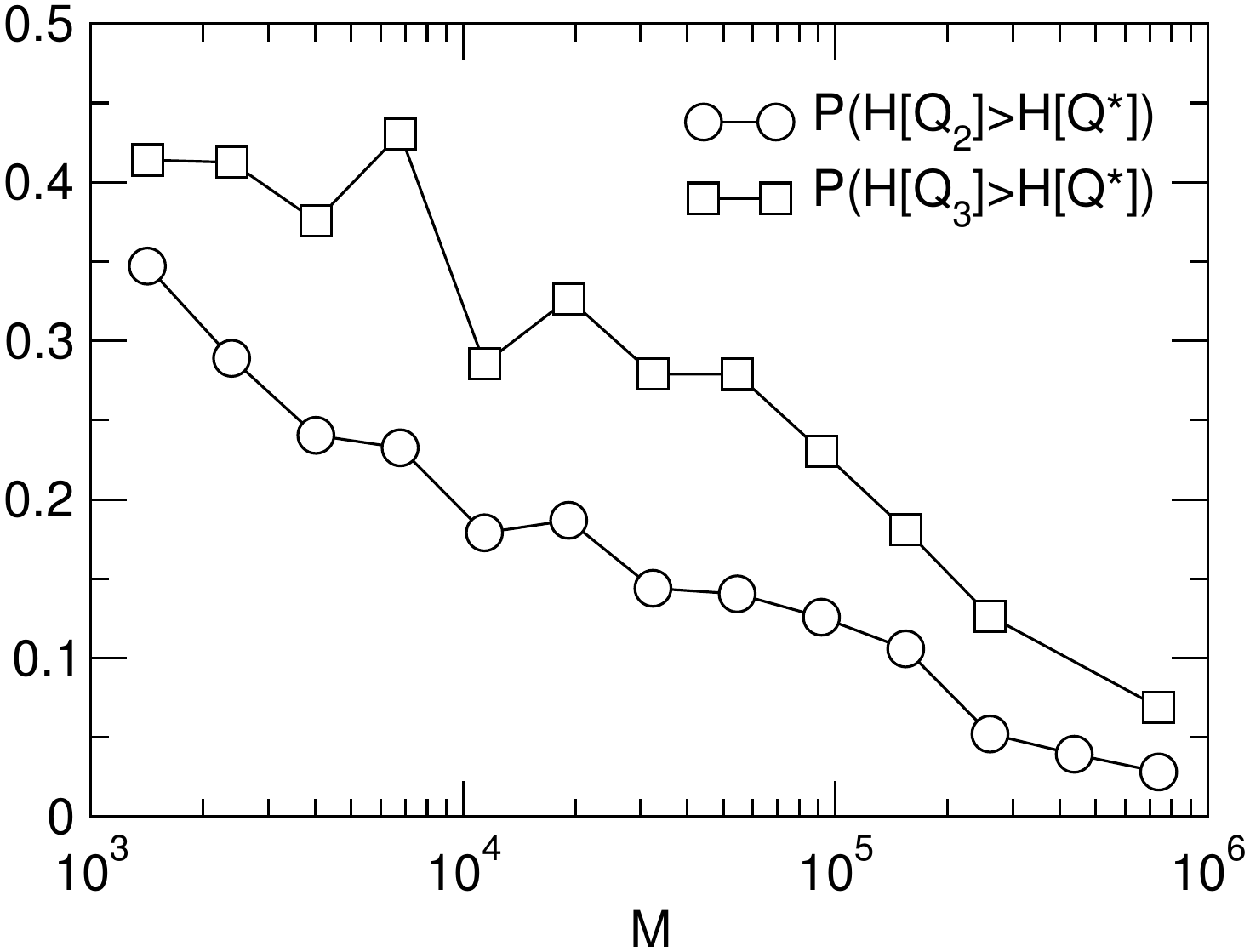}
\caption{Relative relevance of the optimal partition $\mathcal{Q}^*$ with respect to partitions $\mathcal{Q_2}$ and $\mathcal{Q_3}$ (see main text).} 
\label{p-values}
\end{figure}

To check on the robustness of the algorithm presented in section \ref{Qalgorithm}, we compared the $\mathcal{Q}^*$ partition with the solutions obtained via two variations of the algorithm. The first variation consists on choosing the pair of adjacent subsets to be merged in step (ii) at random, and accept the move if the likelihood increases. We name this solution $\mathcal{Q}_2$. The second variation consists in merging triplets of adjacent subsets, selected at random and accepting the move if the likelihood increases. We call this solution $\mathcal{Q}_3$. We draw 50 samples of size M from a distribution $P(s)\sim s^{-\alpha}$, with $\alpha=1$ and compute the models $\mathcal{Q}_2$ and $\mathcal{Q}_3$ 1000 times for each sample. Figure \ref{p-values} shows the probability of finding a partition $\mathcal{Q}_2$ ($\mathcal{Q}_3$) with larger entropy than $\mathcal{Q}^*$. We see that in both cases this probability goes to zero for large sample sizes, meaning that the $\mathcal{Q}^*$ partition is more \emph{relevant} in that limit. We also computed the overlap between $\mathcal{Q}^*$, $\mathcal{Q}_2$ and $\mathcal{Q}_3$ finding overlaps (measured by the Adjusted Rand Index) over $90\%$ for a wide range of parameters ($\alpha \in [0.5,3]$, $M\in [10^3,10^6]$).

\bigskip

\bibliographystyle{ieeetr}
\bibliography{References}

\end{document}